\documentstyle[11pt,aaspp4,tighten]{article}

\slugcomment{draft of \today}

\begin{document}

\font\twelvei = cmmi10 scaled\magstep1 
       \font\teni = cmmi10 \font\seveni = cmmi7
\font\mbf = cmmib10 scaled\magstep1
       \font\mbfs = cmmib10 \font\mbfss = cmmib10 scaled 833
\font\msybf = cmbsy10 scaled\magstep1
       \font\msybfs = cmbsy10 \font\msybfss = cmbsy10 scaled 833
\textfont1 = \twelvei
       \scriptfont1 = \twelvei \scriptscriptfont1 = \teni
       \def\mit{\fam1 }
\textfont9 = \mbf
       \scriptfont9 = \mbfs \scriptscriptfont9 = \mbfss
       \def\bmit{\fam9 }
\textfont10 = \msybf
       \scriptfont10 = \msybfs \scriptscriptfont10 = \msybfss
       \def\bmsy{\fam10 }

\def\etal{{\it et al.~}}
\def\eg{{\it e.g.}}
\def\ie{{\it i.e.}}
\def\lsim{\raise0.3ex\hbox{$<$}\kern-0.75em{\lower0.65ex\hbox{$\sim$}}} 
\def\gsim{\raise0.3ex\hbox{$>$}\kern-0.75em{\lower0.65ex\hbox{$\sim$}}}

\title{The Parker Instability under a Linear Gravity\altaffilmark{8}}
 
\author{Jongsoo Kim\altaffilmark{1,2,5},
        Seung Soo Hong\altaffilmark{2,6},
    and Dongsu Ryu\altaffilmark{3,4,7}}

\altaffiltext{1}
{Korea Astronomy Observatory, San 36-1, Whaam-Dong, Yusong-Ku,
Taejon 305-348, Korea}
\altaffiltext{2}
{Department of Astronomy, Seoul National University, Seoul 151-742, Korea}
\altaffiltext{3}
{Department of Astronomy \& Space Science, Chungnam National University,
Taejon 305-764, Korea}
\altaffiltext{4}
{Department of Astronomy, University of Washington, Box 351580, Seattle,
WA 98195-1580}
\altaffiltext{5}
{e-mail: jskim@hanul.issa.re.kr}
\altaffiltext{6}
{e-mail: sshong@astroism.snu.ac.kr}
\altaffiltext{7}
{e-mail: ryu@sirius.chungnam.ac.kr}
\altaffiltext{8}
{Submitted to the Astrophysical Journal}

\begin{abstract}
A linear stability analysis has been done to a magnetized disk under a
linear gravity.
We have reduced the linearized perturbation equations to a second-order
differential equation which resembles the Schr\"{o}dinger equation with
the potential of a harmonic oscillator.
Depending on the signs of energy and potential terms, eigensolutions can
be classified into ``continuum'' and ``discrete'' families.
When magnetic field is ignored, the continuum family is identified as the
convective mode, while the discrete family as acoustic-gravity waves.
If the effective adiabatic index $\gamma$ is less than unity, the former
develops into the convective instability.
When a magnetic field is included, the continuum and discrete families
further branch into several solutions with different characters.
The continuum family is divided into two modes: one is the original
Parker mode, which is a slow MHD mode modulated by the gravity, and the other
is a stable Alfv\'en mode.
The Parker modes can be either stable or unstable depending on $\gamma$.
When $\gamma$ is smaller than a critical value $\gamma_{\rm cr}$, the
Parker mode becomes unstable.
The discrete family is divided into three modes: a stable fast MHD mode
modulated by the gravity, a stable slow MHD mode modulated by the gravity,
and an unstable mode which is also attributed to a slow MHD mode.
The unstable discrete mode does not always exist.
Even though the unstable discrete mode exists, the Parker mode dominates
it if the Parker mode is unstable.
However, if $\gamma \ge \gamma_{\rm cr}$, the discrete mode could be
the only unstable one.
When $\gamma$ is equal $\gamma_{\rm cr}$, the minimum growth time of the
unstable discrete  mode is $1.3 \times 10^8$ years with a corresponding
length scale of 2.4 kpc.
It is suggestive that the corrugatory features seen in the Galaxy and external
galaxies are related to the unstable discrete mode.
\end{abstract}


\section{INTRODUCTION}

Parker (1966) demonstrated that the equilibrium state of interstellar gas, 
magnetic field, and cosmic-rays under the vertical gravity could
be unstable to long wavelength perturbations along the direction of
the equilibrium magnetic field.
Since then, many authors have studied the details of this instability.
For instance, two-dimensional, nonlinear equilibrium states resulting from
the Parker instability under a uniform gravity were derived by Mouschovias
(1974).
The effects of rotation (Shu 1974; Zweibel \& Kulsrud 1975;
Foglizzo \& Tagger 1994), non-uniform gravity (Horiuchi \etal 1988;
Matsumoto \etal 1988; Giz \& Shu 1993), and skewed magnetic field
(Hanawa \etal 1992) on the Parker instability were discussed.
The Parker instability was considered to be a plausible mechanism
for the formation of large-scale interstellar clouds (Appenzeller 1974;
Mouschovias \etal 1974; Blitz \& Shu 1980; Shibata \& Matsumoto 1991; 
Gomez de Castro \& Pudritz 1992), a possible process for dynamo in
accretion disks (Tout \& Pringle 1992), and the force driving
emergent flux tubes which are closely related to the active phenomena
of the Sun (Shibata \etal 1989a, 1989b; Kaisig \etal 1990; Nozawa \etal 1992).

The original analysis of the instability by Parker (1966) was performed
under a uniform gravity, although he commented on the effects
of a non-uniform, linear gravity.
Since the uniform gravity has a jump at the mid-plane, the physical
quantities should have a cusp.
Because of this, the perturbations in the upper hemisphere cannot
communicate with those in the lower hemisphere.
So Parker's analysis should have been restricted to the ``symmetric''
mode where the vertical velocity perturbation vanishes at the mid-plane.
In order to remove this difficulty, a non-uniform gravity which changes
smoothly at the mid-plane should be introduced.

Giz \& Shu (1993) considered the Parker instability under a gravity whose
functional form is $-g_o\tanh(z/2H_*)$, where $g_o$ and $H_*$ are constants,
and $z$ is the vertical distance from the mid-plane.
With the scale height of stars for $H_*$, this gravity closely represents the
Galactic gravitational field.
In the limit $|z| \rightarrow \infty$ it is reduced to a uniform gravity,
while in $|z| \rightarrow 0$ it is reduced to a linear gravity.
With this gravity, Giz \& Shu obtained a dispersion relation which
matches that under the uniform gravity.
They called it the ``continuum'' family.
But, in addition, they found a dispersion relation which has no
correspondence in the uniform gravity.
They called the latter the ``discrete'' family.
Before Giz \& Shu, Horiuchi \etal (1988) considered the Parker instability
under a point-mass-dominated gravity, which is continuous across the
mid-plane.
It is a good model for the gravitational field in accretion disks.
They considered isothermal perturbations with the effective adiabatic
index $\gamma=1$ and obtained dispersion relations for both the symmetric
and antisymmetric modes.
Their solutions correspond to the continuum family by the Giz \& Shu
classification.
They did not see the solutions of the discrete family, since their
analysis was restricted to isothermal perturbations.

In this paper, we consider a linear gravity whose functional form
is $-g'z$, where $g'$ is a constant.
The linear gravity represents fairly well the $z$-dependence of the
Galactic gravity up to 500 pc (Oort 1965), as well as the ``vertical''
gravity of accretion disks.
It is also easier to handle than those introduced by Giz \& Shu (1993)
and Horiuchi \etal (1988).
Although Parker (1966) considered a linear gravity, all he did was
to show that the equilibrium state of an interstellar gas-field system
becomes also unstable under it.
Here, we investigate the resulting eigensolutions in detail:
we classify them into the continuum and discrete families, discuss their
nature, and give criteria for the existence of the unstable continuum and
discrete families.
Although the existence of the discrete family was already discussed by
Giz \& Shu (1993), our work would serve to get a better insight to its nature
and its possible roles in astrophysics.

Several authors have reported that there exist corrugations in our Galaxy 
(Quiroga \& Schlosser 1977; Lockman 1977; Spicker \& Feitzinger 1986; Alfaro
\etal 1992; Malhotra 1994) and external galaxies (Florido \etal 1991) as well.
Nelson (1976) suggested that the corrugations may be a manifestation of
hydrodynamic acoustic-gravity waves.
The difficulties with Nelson's suggestion are that the characteristic
scale does not come out naturally and that the structures formed by the
acoustic-gravity waves should be oscillatory and transient.
Instead, in this paper we suggest that the corrugations may be formed as
the results of the unstable discrete mode.

The plan of the paper is as follows.
In \S 2 the classification of the eigensolutions into the continuum and 
discrete families is given.
In \S 3 the dispersion relations of the eigensolutions are analyzed in
details.
We first consider the dispersion relations of non-magnetized
disks and then move to those of magnetized disks. 
In \S 4 a summary and discussion are given.
The derivation of eigenvalue equations and the method we used to obtain
the dispersion relation of the continuum family are described in appendices.
  
\section{CONTINUUM AND DISCRETE FAMILIES}

We consider the stability of a disk composed of gas, magnetic field,
and cosmic-rays under a linear gravity.
For the unperturbed equilibrium state, we follow Parker (1966) to assume
that the gas is isothermal
and that the ratio of magnetic pressure to gas pressure, $\alpha$, and the
ratio of cosmic-ray pressure to gas pressure, $\beta$, are spatially
constant.
Then, the unperturbed gas density, gas pressure, cosmic-ray pressure, and
magnetic pressure are given by a gaussian distribution in $z$.
The equations for the stability analysis are obtained by introducing
an infinitesimal perturbation to the unperturbed equilibrium state.
Those equations can be combined into a single eigenvalue equation, which is
a second-order differential equation.
The detailed manipulation of the equations is described in Appendix A.
The resulting eigenvalue equation is
\begin{equation}
\frac{d^2\psi}{d\zeta^2} + (E-V_o\zeta^2)\psi = 0,
\label{master}
\end{equation}
where $E$ and $V_o$ are defined by
\begin{equation}
E \equiv \frac{1}{2} + \frac{\Omega^4 -\left[(1+\alpha+\beta)
+(2\alpha+\gamma)(\nu_x^2+\nu_y^2)\right]\Omega^2+2\alpha\gamma
(\nu_x^2+\nu_y^2)\nu_y^2}{(2\alpha+\gamma)\Omega^2-2\alpha\gamma\nu_y^2},
\eqnum{\ref{E}}
\end{equation}
\begin{equation}
V_o \equiv \frac{1}{4} - \frac{2\alpha(1+\alpha+\beta)\nu_x^2\Omega^2}
{(\Omega^2-2\alpha\nu_y^2)\left[(2\alpha+\gamma)\Omega^2
-2\alpha\gamma\nu_y^2\right]} + \frac{(1+\alpha+\beta)
(1+\alpha+\beta-\gamma)(\nu_x^2+\nu_y^2)} {(2\alpha+\gamma)
\Omega^2-2\alpha\gamma\nu_y^2}.
\eqnum{\ref{V_o}}
\end{equation}
Here $\zeta$ is a normalized $z$-coordinate, $\nu_x$ and $\nu_y$ are
normalized radial and azimuthal wavenumbers, $\Omega$ is a normalized
angular frequency, and  
$\psi$ is a quantity proportional to the normalized vertical velocity 
multiplied by the square root of the unperturbed normalized gas density 
(see Eqs.~[\ref{vzdefine}] and [\ref{norm_vari}]).
It should be pointed out that $\psi^2$ is the kinetic energy density 
associated with the motion in the $z$-direction.
Normalization units are 
the disk scale height $H$, sound speed of unperturbed state $a_s$, 
and unperturbed gas density at the mid-plane $\rho_o(0)$.
Due to the simple functional form of the gravity we used, we could get 
an eigenvalue equation which resembles the Schr\"{o}dinger equation
with the potential of a harmonic oscillator.
As the boundary condition, we assume that $\psi$ stays finite as $\zeta
\rightarrow\infty$.
That is, the energy density perturbation does not diverge at infinite.

The solutions of the eigenvalue equation can be classified into four
different types by the signs of $E$ and $V_o$ (Figure~1).
If $E>0$ and $V_o<0$ so that $E$ is greater than $V(\zeta)$ in the whole
range of $\zeta$, $-\infty < \zeta < +\infty$, all the positive values
are allowed as the eigenvalue $E$  (Type I).
In this case, the eigenfunction $\psi$ oscillates as $|\zeta|$ increases.
If $E<0$ and $V_o<0$ so that $E$ is smaller than $V(\zeta)$ in a finite
interval which includes $\zeta=0$ and greater than $V(\zeta)$ in the rest,
all the negative values are allowed as the eigenvalue (Type II).
The eigenfunction declines exponentially in the finite interval but stays
oscillatory outside it.  
If $E>0$ and $V_o>0$ so that $E$ is greater than $V(\zeta)$ in a finite
interval but smaller than $V(\zeta)$ in the rest, only discrete positive
values are allowed as the eigenvalue (Type III).
The eigenfunction oscillates in the interval but declines exponentially
outside it.
If $E<0$ and $V_o>0$ so that $E$ is smaller than $V(\zeta)$ in the whole
interval, there are no physically meaningful solutions satisfying the
boundary condition (Type IV).
With the tangent hyperbolic gravity, Giz \& Shu (1993) also obtained an
equation similar to Eq.~(\ref{master}) and classified its solutions
into the continuum and discrete families.
According to the Giz \& Shu's classification, the solutions of Type I and II
correspond to the continuum family and  those of Type III to the discrete
family.

\subsection{Continuum Family}

For the continuum family with negative $V_o$, we introduce a new
variable $\xi$ defined by $\xi = \sqrt{2}~^4\!\!\!\sqrt{-V_o}\zeta$ and
transform Eq.~(\ref{master}) into that for the parabolic
cylinder function (Abramowitz \& Stegun 1970):
\begin{equation}
\frac{d^2\psi}{d\xi^2} + \left( \frac{1}{4}\xi^2 - a \right) \psi = 0,
\label{pcf}
\end{equation}
\begin{equation}
a \equiv \frac{1}{2}\frac{-E}{\sqrt{-V_o}}.
\label{c_dispersion}
\end{equation}
As $|\xi|$ increases, the parabolic cylinder function oscillates in
the interval between nodal points.
For $|\xi|\rightarrow\infty$, $\psi \propto \exp(\pm i\xi^2)/\sqrt{\xi}$.
So the energy density perturbation of the continuum family decreases as
$1/|\xi|$ at large $|\xi|$.
It is interesting to notice that, for $|\xi|\rightarrow\infty$, the
$z$ wavelength decreases as $1/|\xi|$, hence the group velocity along the
$z$-direction increases as $|\xi|$.
Consequently the energy flux of the continuum family stays constant
at large $|\xi|$.

The dispersion relations for the symmetric mode can
be calculated by considering the lower boundary condition at the
mid-plane, $\psi(\xi=0)=0$, while the dispersion relations for 
the antisymmetric mode by taking $d\psi(\xi=0)/d\xi=0$ (see Appendix B).
With the condition $\psi\sqrt{\xi}=$ finite at
$|\xi|\rightarrow\infty$, there exists, for a given pair of wavenumbers
$\nu_x^2$ and $\nu_y^2$, an interval of $\Omega^2$ values.
Here, we find solutions with the energy density perturbation zero 
at the upper boundary which is identified to the first nodal point from
the mid-plane, $\xi_{\rm node}$.
Then, the value of $\Omega^2$ is uniquely determined for the given nodal
point.
Since the nodal point indicates the position where the
vertical velocity goes to zero, the distance to this nodal point may be
considered as an ``effective'' wavelength along the $z$-direction.
A method to find the value of $\Omega^2$ with a chosen nodal point
is described in Appendix B.

\subsection{Discrete Family}

For the discrete family with positive $V_o$ and positive $E$, 
Eq.~(\ref{master}) has exactly the same form as the Schr\"{o}dinger
equation with the potential of a harmonic oscillator.  
The solution of the harmonic oscillator problem is well known: the
eigenfunctions are
\begin{equation}
\psi_n(\zeta) = \frac{^8\!\!\!\sqrt{V_o}}{^4\!\!\!\sqrt{\pi}\sqrt{2^n n!}}
H_n \left( ^4\!\!\!\sqrt{V_o}\zeta \right) 
\exp \left( - \frac{\sqrt{V_o}}{2}\zeta^2 \right),
\label{d_eigenfunction}
\end{equation}
and the corresponding eigenvalues are
\begin{equation}
\frac{E}{\sqrt{V_o}} = 2n + 1,
\label{d_dispersion}
\end{equation}
where $H_n$ is the Hermite polynomial of degree $n$, and $n$ takes values
$0,~1,~2, \cdots$.
In this case, $\psi$ decreases exponentially at large $|\zeta|$, so the
energy density perturbation of the discrete family does as well.

The eigenfunctions for $n=1,3, \cdots$ which have $\psi=0$ at $\zeta=0$
represent the symmetric mode, while the eigenfunctions for
$n=0,2, \cdots$ which have $d\psi/d\zeta=0$ at $\zeta=0$
represent the antisymmetric mode.
The eigenvalue relation of Eq.~(\ref{d_dispersion}) gives the dispersion
relation, \ie, $\Omega^2$ as a function of $\nu_x^2$, $\nu_y^2$ and $n$.
Here, $n$ indicates the number of nodal points along the $z$-direction,
so it may be considered as the ``effective'' wavenumber in the $z$-direction.

\section{WAVES AND INSTABILITIES}

In this section, we explore the nature of the continuum and discrete
families in detail by analyzing both the stable (waves) modes and the
unstable (instability) modes.
We first consider the case of non-magnetized disks, then include
a magnetic field and cosmic-rays.

\subsection{Non-Magnetized Disks}

Without a magnetic field, the cosmic-rays are ignored, so
$\alpha=\beta=0$.
Without a magnetic field, there is no preferred direction in the
$x\!\!-\!\!y$ plane, and we can set $\nu_x=0$ without losing any generality. 
Then, $E$ and $V_o$ become  
\begin{equation}
E = \frac{1}{2} + \frac{\Omega^2 - (1+\gamma\nu_y^2)}{\gamma}, 
\label{HD_E}
\end{equation}
\begin{equation}
V_o = \frac{1}{4} + \frac{(1-\gamma)\nu_y^2}{\gamma\Omega^2}.
\label{HD_V_o}
\end{equation}
Eq.~(\ref{master}) with the above $E$ and $V_o$ is exactly the one
Nelson (1976) derived
to study hydrodynamic acoustic-gravity waves under a linear gravity.

In the case of $\gamma=1$ for isothermal perturbations, $V_o=1/4$ 
is always positive.
So, Eq.~(\ref{master}) has only two types of solutions, Type III and
Type IV, depending on the sign of $E$.
The locus of $E=0$ is drawn with a dashed line in the phase
space of $\nu_y^2$ and $\Omega^2$ in Figure~2a.
We are interested only in the upper region of the line with $E>0$ where
the solutions of Type III exist.
The eigenfunctions of the solutions in this region are  
\begin{equation}
\psi(\zeta) = \frac{1}{^4\!\!\!\sqrt{2\pi}} \frac{1}{\sqrt{2^nn!}}
H_n\left(\frac{\zeta}{\sqrt{2}}\right) \exp\left(- \frac{\zeta^2}{4} \right),
\label{eigenfunction}
\end{equation}
and the corresponding dispersion relations are
\begin{equation}
\Omega^2 = \nu_y^2 + n + 1.
\label{iso_dispersion}
\end{equation}
Eq.~(\ref{iso_dispersion}) reduces to the dispersion relation 
$\Omega^2\simeq\nu_y^2$ 
of sound waves for
$\nu_y^2 \gg n+1$.
So the dispersion relations are of the
acoustic-gravity waves which have the characteristics of sound waves
modulated by the vertical gravity, as was pointed out by Nelson (1976).
In Figure~2b, the dispersion relations of the acoustic-gravity waves with
$n=0,~1, \cdots, 6$ are plotted with solid lines.
In this case, no convective mode exits, so all the modes are stable.

If $\gamma\neq 1$, the signs of both $E$ and $V_o$ can be positive or
negative.
In Figures~3a and 4a, three loci of $E=0$, $V_o=0$, and $\Omega^2=0$ 
are drawn for
$\gamma=0.8$ and $1.3$ with a dashed line ($E=0$), a dotted line
($V_o=0$), and a long-dashed line ($\Omega^2=0$).
The phase space of $\nu_y^2$ and $\Omega^2$ is divided into four regions
by the three loci, which are identified as those of the Type II, III, and
IV solutions.
We are interested only in the regions of the Type II
(continuum family) and the Type III (discrete family) solutions.
In Figures~3b and 4b, we plot the dispersion relations of the discrete family
solutions with $n=0,~1, \cdots, 6$ which are obtained from
Eq.~(\ref{d_dispersion}), and the dispersion relations of the continuum family
solutions with nodal points at $\zeta_{\rm node}=3,~5,~10$ which are
computed by the method described in Appendix B.

The solutions of the discrete family, which are identified as
acoustic-gravity waves, always have positive $\Omega^2$, so they are stable.
On the other hand, the solutions of the continuum family, which can be
identified as the convective mode, are stable if $\gamma > 1$, but
unstable if $\gamma < 1$, developing into the convective instability
(Ryu \& Goodman 1992).
The mode with a larger $\zeta_{\rm node}$ is more unstable,
because the gravity
is a linearly increasing function of $\zeta$.

In the case of a uniform gravity, the square of the maximum  growth rate 
of the convective instability is $\Omega^2=-(1-\gamma)/\gamma$ (Ryu \&
Goodman 1992).
For $\gamma=0.8$, it is $-0.25$.
Except at small wavenumbers, the growth rate of the convective
instability under the linear gravity in Figure~3b is larger than
the maximum growth rate under the uniform gravity.

The dispersion relations of the continuum family solutions shown in 
Figures~3b and 4b are for the symmetric mode.
In order to see the effects of the boundary condition at the mid-plane,
we have also calculated the dispersion relations of the antisymmetric mode.
Figure~5 shows the resulting dispersion relations with nodal points
at $\zeta_{\rm node}=2$ and $3$.  
Solid lines represent the dispersion relations of the symmetric mode, and
dotted lines those of the antisymmetric mode.
The growth rates of the both modes are nearly the same, unless
$\zeta_{\rm node}\ll 1$.  
This is true, even if the magnetic field is included.
The reason for this is that the gravity is small near the mid-plane.
So the details around the mid-plane are not important unless
$\zeta_{\rm node}\ll 1$.
In the rest of the paper, we then present only the dispersion relation of
the symmetric mode for the continuum family solutions.

\subsection{Magnetized Disks}

\subsubsection{Limit of $H \rightarrow \infty$}

Taking the limit $H \rightarrow \infty$ means also making
$g' \rightarrow 0$.
So the problem reduces to that of finding the dispersion relation in
a uniform medium without gravity.
Since the perturbation of cosmic-ray pressure is zero in the uniform
medium (see [\ref{o_cosmic}]), we should get the dispersion relations for three
magnetohydrodynamic (MHD) waves.
From (\ref{o_continuity}) to (\ref{o_energy}), by setting $d/dz=0$, we have
\begin{equation}
\left(\frac{\omega^2}{a_s^2}-2\alpha k_y^2\right)
\left[\frac{\omega^4}{a_s^4}-(2\alpha+\gamma)(k_x^2+k_y^2)
\frac{\omega^2}{a_s^2}
+2\alpha\gamma(k_x^2+k_y^2)k_y^2 \right]\Psi = 0,
\end{equation}
where $a_s$ is the isothermal sound speed.
The above dispersion relation can be decoupled into one for
the Alfv\'en waves
\begin{equation}
\omega^2 - v_A^2 k_y^2 = 0,
\label{disp_alf}
\end{equation}
and the other for the slow and the fast MHD waves
\begin{equation}
\omega^4 - (v_A^2+c_s^2)(k_x^2+k_y^2)\omega^2 +
v_A^2 c_s^2 k_y^2(k_x^2+k_y^2)=0.
\label{disp_fs}
\end{equation}
Here, $v_A = 2\alpha a_s^2$ and $c_s^2 = \gamma a_s^2$ are the Alfv\'en
and adiabatic sound speeds.

\subsubsection{Case of $\nu_x = 0$}

It is known that the growth rate of the Parker instability is larger
for perturbations with larger $x$ wavenumbers (Parker 1967).
However, in this subsection we first consider the case of $\nu_x=0$
in detail, since the equations involved are much simple and yet most
important physical aspects are included.
We comment briefly on the general case of $\nu_x\ne 0$ in the next
subsection.

In the case of $\nu_x=0$, the $x$-components of velocity and
magnetic field perturbations are decoupled from the rest and result in
Alfv\'en waves with the dispersion relation (\ref{disp_alf})
(see Eqs.~[\ref{o_vx}] and [\ref{o_bx}]).
So, with the rest of the perturbation equations, we get the eigenvalue
equation (\ref{master}) with $E$ and $V_o$ given by
\begin{equation}
E = \frac{1}{2} 
+\frac{\Omega^4-[(1+\alpha+\beta)+(2\alpha+\gamma)\nu_y^2]\Omega^2+2
\alpha\gamma\nu_y^4} {(2\alpha+\gamma)\Omega^2-2\alpha\gamma\nu_y^2},
\label{e_for_vx0}
\end{equation}
\begin{equation}
V_o = \frac{1}{4}+\frac{(1+\alpha+\beta)(1+\alpha+\beta-\gamma)\nu_y^2}
{(2\alpha+\gamma)\Omega^2-2\alpha\gamma\nu_y^2}.
\label{v_for_vx0}
\end{equation}

In order to classify its solutions, we draw in Figure~6a the loci of $E=0$,
$V_o=0$, and $(2\alpha+\gamma)\Omega^2-2\alpha\gamma\nu_y^2=0$ with two
dashed lines, a dotted line, and a long-dashed line, respectively, for
$\alpha=1$, $\beta=0$, and $\gamma=1$.
Since $E=0$ reduces to a quadratic equation of $\Omega^2$, there exist 
two branches, which are labeled by $E^+=0$ (the positive branch) and $E^-=0$
(the negative branch).
The loci of $E^-=0$ and $(2\alpha+\gamma)\Omega^2-2\alpha\gamma\nu_y^2 = 0$
intersect at the point where $\nu_y^2=(2\alpha+\gamma)(1+\alpha+\beta)
/(2\alpha\gamma)$ and $\Omega^2=1+\alpha+\beta$.
The intersection point exists also in the  case of $\nu_x^2 \ne 0$, and
its position is determined only by $\alpha$, $\beta$, and $\gamma$.
So, its location for the case of $\nu_x^2 \ne 0$ is the same as that for
the case of $\nu_x^2 = 0$.
The four loci divide the phase space of $\nu_y^2$ and $\Omega^2$ into
six regions, which are identified in Figure~6a by the type of solutions.
The continuum family solutions exist in the regions of Type I and II,
and the discrete family solutions in Type III.
Figure~6b shows the dispersion relations of the discrete family solutions
with $n=0,~1,~2$ and those of the continuum family solutions with
nodal points at $\zeta_{\rm node}=3,~5,~10$.

The continuum family solutions correspond to those of the original Parker
instability and are identified as a slow MHD mode modulated by the gravity
(Parker 1966; Shu 1974).
For the same nodal point and the same parameters ($\alpha$, $\beta$,
$\gamma$), the growth rate of the continuum family is larger under
a linear gravity than under a uniform gravity.
A full comparative study of the growth rate under different models of
gravity will be given in a separate paper.

The discrete family solutions in the upper left region of Figure~6b are
MHD counterparts of Nelson's acoustic-gravity waves, which are
identified as fast MHD waves modulated by gravity.
The discrete family solutions in the region enclosed by the loci of $E^-=0$
and $(2\alpha+\gamma)\Omega^2-2\alpha\gamma\nu_y^2 = 0$ represent a new mode
which was not discovered in any previous work.
They are always stable, since $\Omega^2>0$ at the intersection point
of the loci.
On the following grounds, we identify them as slow MHD waves
modulated by gravity.
For $\nu_y\rightarrow\infty$ and $n\rightarrow\infty$, the dispersion
relation of the discrete family becomes
\begin{equation}
\Omega^4-(2\alpha+\gamma)(\nu_y^2+\nu_z^2)\Omega^2+2\alpha\gamma
(\nu_y^2+\nu_z^2)\nu_y^2=0,
\label{disc_disp_for_vx0}
\end{equation}
if we set $2\sqrt{V_o}n=\nu_z^2$ (see Eqs.~[\ref{d_dispersion}],
[\ref{e_for_vx0}], and [\ref{v_for_vx0}]).
Note that $V_o$ becomes constant in this limit.
Eq.~(15) is the dispersion relation for the fast and slow waves
(see Eq.~[\ref{disp_fs}]).
The fast waves correspond to the solutions in the upper left region, while
the slow waves to those in the region enclosed by loci of $E^-=0$ and
$(2\alpha+\gamma)\Omega^2-2\alpha\gamma\nu_y^2 = 0$.

If the slope of the locus of $V_o=0$ in the phase space of $\nu_y^2$ and
$\Omega^2$ is equal to or greater than zero, the continuum family solutions
are stabilized.
For given $\alpha$ and $\beta$, the critical value of the adiabatic index,
$\gamma_{\rm cr}$, which gives zero slope, is
\begin{equation}
\gamma_{\rm cr} = \frac{(1+\alpha+\beta)^2}{1+\frac{3}{2}\alpha+\beta}.
\end{equation}
Note that this is same as the critical value for the case of a uniform gravity
(Parker 1966).
To see the nature of the solutions in the case of $\gamma\geq\gamma_{\rm cr}$,
we draw in Figure~7a the loci of $E=0$, $V_o=0$, and
$(2\alpha+\gamma)\Omega^2-2\alpha\gamma\nu_y^2=0$
for $\alpha=1$, $\beta=0$, and $\gamma=\gamma_{\rm cr}$.
Figure~7b presents the dispersion relations of the discrete family
solutions with $n=0,~1,~2$ and those of the continuum family solutions
with nodal points at $\zeta_{\rm node}=3,~5,~10$.
As in the case of Figure~6, there are solutions of the continuum family
in the regions of Type I and II, but now they are stable with
$\Omega^2\geq0$.
Also there are solutions of two stable discrete modes in the
upper left region and in the region enclosed by loci of $E^-=0$ and
$(2\alpha+\gamma)\Omega^2-2\alpha\gamma\nu_y^2 = 0$.
However, there is another region of Type III enclosed by loci
of $E^-=0$ and $V_o=0$, as can be seen in the enlarged plot of Figure~7c.
The solutions of the discrete mode in this region are unstable with
$\Omega^2<0$.
There are an infinite number of unstable solutions with $n=0,~1,~2,\cdots$,
and Figure~7d shows the dispersion relations with $n=0,~1,~2$.

Figure~8 shows the eigenfunctions of the unstable discrete mode solutions
with $n=0,~1,~2$ for $\alpha=1$, $\beta=0$, $\gamma=\gamma_{\rm cr}$,
and $\nu_x=0$.  For a given $n$, the most unstable growth rate
and its azimuthal wavenumber are used.
The eigenfunctions of odd $n$'s are of the symmetry mode with $\psi(0)=0$,
while those of even $n$'s the antisymmetry mode with
$d\psi(0)/d\zeta=0$.
As was noted in \S 2.2, the energy density perturbation of the discrete mode
solutions is concentrated around $\zeta=0$.
Since the stable discrete mode solutions
in the upper left region are identified as the fast MHD mode modulated by
gravity, we may attribute the unstable discrete solutions to the slow MHD
mode modulated by gravity as the stable discrete solutions in the region
enclosed by loci of $E^-=0$ and
$(2\alpha+\gamma)\Omega^2-2\alpha\gamma\nu_y^2 = 0$.

The unstable discrete mode does not always exist.
Its existence depends on whether the two loci of $E^-=0$ and $V_o=0$ meet
at two points: one at the origin $(\nu_y^2=0, \Omega^2=0)$ and the
other denoted by $(\nu_{y,\rm int}^2,~\Omega_{\rm int}^2)$.
For $\beta=0$, Figure~9a shows the equi-$\nu_{y,\rm int}^2$ contours on
the $[\alpha, \gamma]$ plane, and Figure~9b does the
equi-$\Omega_{\rm int}^2$ contours.
The unstable discrete mode exists if $\gamma$ has a value in the region
between the two lines of $\Omega_{\rm int}^2=0$ in Figure~9b, or if
$1<\gamma<\gamma_{\rm int}$.
Here, $\gamma_{\rm int}$ is a function of $\alpha$ in this case of $\beta=0$
and $\nu_x=0$, but it depends on other parameters, too, in general.
If a medium has $1<\gamma<\min(\gamma_{\rm cr},~\gamma_{\rm int})$, both
the unstable continuum and discrete modes exist.
The growth rate of the unstable continuum mode, especially at small
wavelengths, is always larger than that of the unstable discrete mode,
making the continuum mode dominate over the discrete mode.
On the other hand, if $\gamma_{\rm cr}\leq\gamma<\gamma_{\rm int}$,
the discrete mode is the only unstable one, so it determines the evolution
of the medium.

If the scale height (160 pc) and the velocity dispersion (6.4 km/sec) of 
interstellar clouds are substituted for $H$ and $a_s$, respectively,
the growth time of the most unstable discrete solution with $n=0$ becomes
$1.3\times 10^8$ years at the wavelength of $2.4$ kpc.

\subsubsection{Case of $\nu_x \neq 0$}

When $\nu_x \neq 0$, the Alfv\'en mode is no longer decoupled from the
other modes.
Figure~10a presents
the loci of $E=0$, $V_o=0$, $(2\alpha+\gamma)\Omega^2-2\alpha\gamma\nu_y^2=0$,
and $\Omega^2-2\alpha\nu_y^2=0$ for $\alpha=1$, $\beta=0$, $\gamma=1$, and
$\nu_x=1$.
In this general case, both loci of $E=0$ and 
$V_o=0$ are divided into positive and negative branches. 
The locus of $\Omega^2-2\alpha\nu_y^2=0$ is drawn with a dot-dashed line.
The six loci divide the phase space of $\nu_y^2$ and $\Omega^2$
into eight regions.
Since $\gamma=1$ there is no region containing the unstable discrete
mode solutions.
The solutions of the continuum and discrete families identified in Figures~6
and 7 are also found in the figure,
and Figure~10b shows their dispersion relations.
In addition, we have a region of Type II enclosed by the loci of $V_o^+=0$
and $\Omega^2-2\alpha\nu_y^2=0$.
The solutions in this region are of the continuum family and always stable.
Since one of the boundaries of this region corresponds to the dispersion
relation of the Alfv\'en waves, we attribute its solutions to the
Alfv\'en mode.

\section{SUMMARY AND DISCUSSION}

We have performed a linear stability analysis for magnetized disks under 
a linear gravity.
The linearized perturbation equations have been reduced to an eigenvalue
equation that has the same form as the Schr\"{o}dinger equation with an
energy $E$ and a potential $V_o\zeta^2$, where $E$ and $V_o$ are give by
Eqs.~(\ref{E}) and (\ref{V_o}). 
By the signs of $E$ and $V_o$, the eigensolutions have been
classified into two families:
the continuum family for $V_o < 0$ (independent of the sign of $E$) and the
discrete family for $E>0$ and $V_o>0$.

Without a magnetic field, 
the solutions of the continuum family are identified as the
convective mode and those of the discrete family as the acoustic-gravity
waves which were previously discussed by Nelson (1976).
The convective mode develops into an instability if $\gamma < 1$ (Ryu
\& Goodman 1992).

With a magnetic field, the solutions of the continuum family are divided
into two modes: a stable Alfv\'en mode, and 
the original Parker mode.  The Parker mode is a slow MHD mode modulated
by the gravity, 
and becomes unstable when the effective adiabatic index
is smaller than the critical value $\gamma_{\rm cr}=
(1+\alpha+\beta)^2/(1+3\alpha/2+\beta)$ for $\nu_x=0$.
The solutions of the discrete family are divided into three modes:
a stable fast MHD mode modulated by the gravity, a stable slow MHD mode
modulated by the gravity, and an unstable mode.
The two stable discrete modes always exist, but the unstable discrete mode
exists only if $1<\gamma<\gamma_{\rm int}$, where $\gamma_{\rm int}$ is a
function of the parameters of the problem.
The unstable discrete mode was first discovered by Giz \& Shu (1993).

In the Galactic context, we discuss the possible roles of the unstable
modes of the continuum and discrete solutions on the formation of giant 
molecular clouds and corrugations.  For the scale height and
the velocity dispersion of the cloud distribution, we take 160 pc and 
6.4 km/sec (Falgarone \& Lequeux 1973).  
Then, the maximum growth time and the corresponding length of the continuum 
solution are $1.5\times 10^7$ years and 340 pc, for the parameters of 
$\alpha=0.25, \beta=0.4, \gamma=1.0, \nu_x=0.0$ and $\zeta_{\rm node}=3$.  
Here, the values of $\alpha$ and $\beta$ are the canonical ones of 
the interstellar medium (Spitzer 1978) and the chosen nodal point
is $\sim 500$ pc (Elmegreen 1982).  
On the base of the time and length scales of the Parker instability 
under the uniform gravity, Mouschovias {\it et al.} (1974) preferred
the interstellar medium behind Galactic shocks to the canonical interstellar
medium since the growth time of the Parker instability with the canonical
parameters of the interstellar medium is larger than the life time of the 
giant molecular clouds, $3\times10^7$ years (Blitz \& Shu 1980).  
Under the linear gravity, the unstable modes of the continuum solution
may, however, develop within the life time.  Therefore, developments
of the unstable modes of the continuum solution of the Parker instability
under the linear gravity are not confined to spiral arms but ubiquitous
in the Galaxy.

We also suggest that the corrugations observed in our Galaxy and external
galaxies may have been formed by the unstable discrete mode.
For parameters, $\alpha=1$, $\beta=0$, $\gamma=\gamma_{\rm cr}$, and
$\nu_x=0$, the growth time of the most unstable discrete solution with
$n=0$ is $1.3\times 10^8$ years at the wavelength of $2.4$ kpc.
This time scale is comfortably shorter than the age of the galaxies, and
the length scale is consistent with the characteristic wavelength of the
vertical structure of the Carina-Sagittarius spiral arm (Alfaro \etal 1992).
Furthermore, the antisymmetric nature of the $n=0$ solution is also 
consistent with
the corrugations found by Alfaro \etal (1992).

In this paper, we have neglected the effects of rotation on the Parker
instability.
Although the effects of differential rotation which occurs in
accretion disks as well as galactic disks are difficult to work with
analytically, those of rigid-body rotation can be included in the frame
of the present work.
The inclusion of rigid-body rotation is expected to have two major
effects on our results.
First, although it does not change the criterion of the instability,
the inclusion of rigid-body rotation reduces the growth rate as already
pointed in Shu (1974) and Zweibel \& Kulsrud (1975).
Second, also as pointed in Shu (1974), with rigid-body rotation the Lindblad
oscillation enters to the problem making the classification, for instance
the one in Figure 10, more complicated.
We are working on these and the results will be reported in a separate paper.

\acknowledgments
At an early stage of this work JK was supported by DAEWOO 
Post-Graduate Scholarship.
The work by SSH and JK was supported 
by Seoul National University DAEWOO Research Fund.  
The work by DR was supported in part by Seoam Scholarship Foundation.
We are grateful to Dr.~T.~W.~Jones and the referee, Dr.~E.~G.~Zweibel,
for comments on the manuscript.

\appendix

\section{EQUILIBRIUM STATES AND LINEARIZED PERTURBATION EQUATIONS}

The MHD equations for gas with cosmic-rays are given by
\begin{equation}
\frac{\partial \rho}{\partial t} + {\bmsy \nabla} \cdot (\rho {\bmit v} )
= 0,
\label{continuity}
\end{equation}
\begin{equation}
\rho \left( \frac{\partial {\bmit v}}{\partial t} + {\bmit v} \cdot
{\bmsy \nabla} {\bmit v} \right) = -{\bmsy \nabla} p - {\bmsy \nabla}
p_{\rm cr} - {\bmsy \nabla} \frac{B^2}{8\pi} + \frac{1}{4\pi} {\bmit B}
\cdot {\bmsy \nabla} {\bmit B} + \rho {\bmit g},
\label{momentum}
\end{equation}
\begin{equation}
\frac{\partial {\bmit B}}{\partial t} = {\bmsy \nabla} \times
( {\bmit v} \times {\bmit B}),
\label{faraday}
\end{equation}
\begin{equation}
\left( \frac{\partial}{\partial t} + {\bmit v} \cdot {\bmsy \nabla}
\right) \left( \frac{p}{\rho^\gamma} \right) = 0,
\label{energy}
\end{equation}
\begin{equation}
{\bmit B} \cdot {\bmsy \nabla} p_{\rm cr} = 0,
\label{cosmic-ray}
\end{equation}
where $p_{\rm cr}$ is the cosmic ray pressure.

For cosmic-rays we use the equation which was employed in the study of the
effect of rotation on the Parker instability by Shu (1974).
It is valid on the assumption that cosmic-rays adjust themselves to
flows with an infinite speed along magnetic field lines.
If the detail dynamics of cosmic-rays is important, for instance, around
the shocks where they are accelerated, it can be treated either by the
diffusion-advection equation such as derived by Skilling (1975) or by the
two-fluid model (Drury \& V\"olk 1981).
In the two-fluid model, cosmic-rays are represented by a fluid with an
adiabatic index between 4/3 and 5/3.
With these more realistic treatments of cosmic-rays, it was shown that
fluids with cosmic-rays are subject to cosmic-ray induced instabilities
with a scale comparable to the diffusion length of cosmic-rays (see,
\eg~Drury \& Falle 1986; Zank, Axford, \& McKenzie 1990; Kang, Jones,
\& Ryu 1992; Ryu, Kang, \& Jones 1993; Begelman \& Zweibel 1994).
However, since here we are interested in the Parker instability which occurs
in a much larger scale mainly by the coupling of magnetic field with gravity,
the above simpler equation for cosmic-rays should be enough.

To describe the local behavior of the Parker instability in magnetized
disks, we introduce Cartesian coordinates $(x,~y,~z)$ whose directions are
radial, azimuthal and vertical, respectively.  
We assume that the gravity, $\bmit g$, is vertical and a linear function
of $z$, ${\bmit g} = (0, 0, - g' z)$,
and that unperturbed magnetic field, ${\bmit B}_o$, is a
function of $z$ and has only an azimuthal component, 
${\bmit B}_o = [0, B_o(z), 0]$. Here, $g'$ is a positive constant,
We further assume that unperturbed gas is isothermal and the ratio of
its magnetic pressure to gas pressure, $\alpha$, and the ratio of its
cosmic-ray pressure to gas pressure, $\beta$, are spatially constant,
\ie, $p_o = a_s^2 \rho_o$, $B_o^2 / (8\pi) = \alpha p_o$, and
$p_{{\rm cr},o} = \beta p_o$, where $a_s$ is the isothermal sound speed,
the quantities with subscript $_o$ indicate unperturbed quantities.
Then, the vertical distribution of unperturbed quantities is
\begin{equation}
\frac{\rho_o(z)}{\rho_o(0)} = \frac{p_o(z)}{p_o(0)} = 
\frac{p_{{\rm cr,}o}(z)}{p_{{\rm cr,}o}(0)} = 
\frac{B_o^2(z)}{B_o^2(0)} =
\exp \left[ - \frac{1}{2} \left( \frac{z}{H} \right)^2 \right], 
\label{equilibrium}
\end{equation}
where $H^2 \equiv (1+\alpha+\beta) a_s^2 / g'$ and $\rho_o(0)$, $p_o(0)$,
$p_{{\rm cr},o}(0)$, $B_o(0)$ are the values of $\rho_o(z)$, $p_o(z)$,
$p_{{\rm cr},o}(z)$, $B_o(z)$ at $z=0$. 

We perturb the above equilibrium state.
The perturbed state is described by
\begin{equation}
{\bmit v} ; \; \rho = \rho_o + \delta\rho; \; p = p_o + \delta p; \;
p_{\rm cr} = p_{{\rm cr},o} + \delta p_{\rm cr}; \; {\bmit B} =
B_o\hat{{\bmit e}}_y + \delta {\bmit B}.
\label{perturbations}
\end{equation}
If we insert the quantities in Eq.~(\ref{perturbations}) into
Eqs.~(\ref{continuity})-(\ref{cosmic-ray}) and keep only the first-order
terms of the perturbed variables, we obtain the linearized perturbation
equations
\begin{equation}
\frac{\partial\delta\rho}{\partial t} - \frac{\rho_o}{H^2} z v_z
+ \rho_o \left( \frac{\partial v_x}{\partial x} +\frac{\partial v_y}
{\partial y} + \frac{\partial v_z}{\partial z} \right) = 0,
\label{p_continuity}
\end{equation}
\begin{equation}
\rho_o\frac{\partial v_x}{\partial t} + \frac{\partial \delta p}
{\partial x} + \frac{\partial \delta p_{\rm cr}}{\partial x}
+ \frac{B_o}{4\pi}\frac{\partial \delta B_y}{\partial x}
- \frac{B_o}{4\pi}\frac{\partial \delta B_x}{\partial y} = 0,
\end{equation}
\begin{equation}
\rho_o \frac{\partial v_y}{\partial t} + \frac{\partial \delta p}
{\partial y} + \frac{\partial \delta p_{\rm cr}}{\partial y}
+ \frac{B_o}{8\pi H^2}z \delta B_z = 0,
\end{equation}
\begin{equation}
\rho_o \frac{\partial v_z}{\partial t} + \frac{\partial \delta p}
{\partial z} + \frac{\partial \delta p_{\rm cr}}{\partial z}
- \frac{B_o}{8\pi H^2}z\delta B_y - \frac{B_o}{4\pi}\frac
{\partial\delta B_z}{\partial y} + \frac{B_o}{4\pi}\frac
{\partial\delta B_y}{\partial z} + \frac{(1+\alpha+\beta)a_s^2}{H^2}
z\delta\rho = 0,
\end{equation}
\begin{equation}
\frac{\partial\delta B_x}{\partial t} - B_o\frac{\partial v_x}
{\partial y} = 0,
\end{equation}
\begin{equation}
\frac{\partial\delta B_y}{\partial t} + B_o\frac{\partial v_x}{\partial x}
- \frac{B_o}{2H^2}zv_z + B_o\frac{\partial v_z}{\partial z} =0,
\end{equation}
\begin{equation}
\frac{\partial\delta B_z}{\partial t} - B_o\frac{\partial v_z}
{\partial y} = 0,
\end{equation}
\begin{equation}
\frac{\partial\delta p}{\partial t} - \frac{p_o}{H^2}zv_z
+ \gamma p_o \left( \frac{\partial v_x}{\partial x} + \frac{\partial v_y}
{\partial y} + \frac{\partial v_z}{\partial z} \right) = 0,
\end{equation}
\begin{equation}
\frac{\partial\delta p_{\rm cr}}{\partial t} - \frac{p_{{\rm cr},o}}
{H^2}zv_z = 0.
\label{p_cosmic}
\end{equation}

Since the coefficients of Eqs.~(\ref{p_continuity})-(\ref{p_cosmic})
do not depend explicitly on $x,~y,~t$, the perturbed quantities can be
Fourier-decomposed with respect to $x,~y,~t$.
So the perturbed quantities have the following form
\begin{equation}
\delta Q(x,y,z;t) = \delta Q(z) \exp (i\omega t -ik_xx - ik_yy),
\label{eigensolution}
\end{equation}
where $\omega$ is the angular frequency and $k_x$ and $k_y$ are the
wavenumbers along the $x$ and $y$ directions.
Inserting the above into Eqs.~(\ref{p_continuity})-(\ref{p_cosmic}), we
get
\begin{equation}
i\omega\delta\rho -\frac{\rho_o}{H^2}zv_z + \rho_o \left(-ik_xv_x
-ik_yv_y+\frac{dv_z}{dz}\right) = 0,
\label{o_continuity}
\end{equation} 
\begin{equation}
i\omega\rho_o v_x -ik_x\delta p -ik_x\delta p_{\rm cr} -ik_x\frac{B_o}
{4\pi}\delta B_y +ik_y\frac{B_o}{4\pi}\delta B_x = 0,
\label{o_vx}
\end{equation}
\begin{equation}
i\omega\rho_o v_y - ik_y\delta p -ik_y\delta p_{\rm cr} + \frac{B_o}
{8\pi H^2}z\delta B_z = 0,
\end{equation}
\begin{equation}
i\omega\rho_o v_z + \frac{d\delta p}{dz} + \frac{d\delta p_{\rm cr}}{dz}
- \frac{B_o}{8\pi H^2}z\delta B_y + ik_y\frac{B_o}{4\pi}\delta B_z 
+ \frac{B_o}{4\pi}\frac{d\delta B_y}{dz} + \frac{(1+\alpha+\beta)a_s^2}
{H^2}z\delta\rho = 0,
\end{equation}
\begin{equation}
i\omega\delta B_x + ik_y B_o v_x = 0,
\label{o_bx}
\end{equation}
\begin{equation}
i\omega\delta B_y -ik_x B_o v_x - \frac{B_o}{2H^2}zv_z +
B_o\frac{dv_z}{dz} = 0,
\end{equation}
\begin{equation}
i\omega\delta B_z + ik_y B_o v_z = 0,
\end{equation}
\begin{equation}
i\omega\delta p - \frac{p_o}{H^2}zv_z + \gamma p_o \left(-ik_x v_x 
-ik_y v_y + \frac{dv_z}{dz} \right) = 0,
\label{o_energy}
\end{equation}
\begin{equation} 
i\omega\delta p_{\rm cr} - \frac{p_{{\rm cr},o}}{H^2}zv_z = 0.
\label{o_cosmic}
\end{equation}

The above nine equations (Eqs.~[\ref{o_continuity}]-[\ref{o_cosmic}]) with
nine unknowns ($\delta\rho$, $v_x$, $v_y$, $v_z$, $\delta B_x$,
$\delta B_y$, $\delta B_z$, $\delta p$, $\delta p_{\rm cr}$) can be 
combined into a single second-order differential equation for one unknown,
\begin{eqnarray}
\lefteqn{
\frac{d^2v_z}{dz^2}
- \frac{z}{H^2}\frac{dv_z}{dz}
+ \frac{ \omega^4/a_s^4
         -\left[ (1+\alpha+\beta)/H^2 
         +(2\alpha+\gamma)(k_x^2+k_y^2)\right]\omega^2/a_s^2
         +2\alpha\gamma(k_x^2+k_y^2)k_y^2 }
       { (2\alpha+\gamma)\omega^2/a_s^2
         -2\alpha\gamma k_y^2 } v_z } \nonumber \\
  && 
+ \left\{
  \frac{2\alpha(1+\alpha+\beta)k_x^2\omega^2/a_s^2}
       {\left(\omega^2/a_s^2-2\alpha k_y^2\right)
        \left[(2\alpha+\gamma)\omega^2/a_s^2
                    - 2\alpha\gamma k_y^2 \right]}
- \frac{(1+\alpha+\beta)(1+\alpha+\beta-\gamma)(k_x^2+k_y^2)}
       {(2\alpha+\gamma)\omega^2/a_s^2-2\alpha\gamma k_y^2}
  \right\} \nonumber \\
  &&
\times  \frac{z^2}{H^4} v_z = 0.
\label{vz_master}
\end{eqnarray}
With $\Psi$ defined as
\begin{equation}
v_z = \Psi \exp\left[\frac{1}{4}\left(\frac{z}{H}\right)^2\right],
\label{vzdefine}
\end{equation}
the equation reduces to a second-order differential equation
without the first-order derivative term
\begin{eqnarray}
\lefteqn{
\frac{d^2\Psi}{dz^2}
+ \left\{\frac{1}{2H^2}
         + \frac{ \omega^4/a_s^4
                  -\left[(1+\alpha+\beta)/H^2
                       +(2\alpha+\gamma)(k_x^2+k_y^2)\right]\omega^2/a_s^2
                  +2\alpha\gamma(k_x^2+k_y^2)k_y^2}
                {(2\alpha+\gamma)\omega^2/a_s^2-2\alpha\gamma k_y^2}
  \right\} \Psi } \nonumber \\
  &&
- \left\{  \frac{1}{4} 
         - \frac{2\alpha(1+\alpha+\beta)k_x^2\omega^2/a_s^2}
                {\left(\omega^2/a_s^2-2\alpha k_y^2\right)
                 \left[(2\alpha+\gamma)\omega^2/a_s^2
                       -2\alpha\gamma k_y^2\right]}
         + \frac{(1+\alpha+\beta)(1+\alpha+\beta-\gamma)(k_x^2+k_y^2)}
                {(2\alpha+\gamma)\omega^2/a_s^2
                 -2\alpha\gamma k_y^2}
  \right\} \nonumber \\
  &&
\times  \frac{z^2}{H^4} \Psi=0.
\label{Psi}
\end{eqnarray} 

We define the dimensionless variables by
\begin{equation}
\Omega \equiv \omega H/a_s, \; \nu_x \equiv k_x H, \; 
\nu_y \equiv k_y H, \; \psi \equiv \Psi/a_s, \; \zeta \equiv z/H.
\label{norm_vari}
\end{equation}
Then, the combined perturbation equation becomes 
\begin{equation}
\frac{d^2\psi}{d\zeta^2} + (E-V_o\zeta^2)\psi = 0,
\label{ap:master}
\end{equation}
where $E$ and $V_o$ are given by
\begin{equation}
E \equiv \frac{1}{2} + \frac{\Omega^4 -\left[(1+\alpha+\beta)
+(2\alpha+\gamma)(\nu_x^2+\nu_y^2)\right]\Omega^2+2\alpha\gamma
(\nu_x^2+\nu_y^2)\nu_y^2}{(2\alpha+\gamma)\Omega^2-2\alpha\gamma\nu_y^2},
\label{E}
\end{equation}
\begin{equation}
V_o \equiv \frac{1}{4} - \frac{2\alpha(1+\alpha+\beta)\nu_x^2\Omega^2}
{(\Omega^2-2\alpha\nu_y^2)\left[(2\alpha+\gamma)\Omega^2
-2\alpha\gamma\nu_y^2\right]} + \frac{(1+\alpha+\beta)
(1+\alpha+\beta-\gamma)(\nu_x^2+\nu_y^2)} {(2\alpha+\gamma)
\Omega^2-2\alpha\gamma\nu_y^2}.
\label{V_o}
\end{equation}
 
\section{METHOD TO COMPUTE THE DISPERSION RELATIONS OF THE CONTINUUM FAMILY 
SOLUTIONS}

For given $\alpha, \beta$, and $\gamma$, $E$ and $V_o$ in Eqs.~(\ref{E})
and (\ref{V_o}) are functions of $\nu_x^2, \nu_y^2$, and $\Omega^2$.
So getting the dispersion relations is equivalent to finding the
values of $\nu_x^2$, $\nu_y^2$, and $\Omega^2$ which satisfy the boundary
conditions imposed on Eq.~(\ref{ap:master}).
We consider the upper boundary condition $\psi = 0$ at $\zeta=
\zeta_{\rm node}$ and the lower boundary condition $\psi(0)=0$ or
$d\psi(0)/d\zeta=0$ at the mid-plane.
Here, $\zeta_{\rm node}$ is the nearest nodal point from the mid-plane.
The condition $\psi(0)=0$ doesn't allow gas to cross the mid-plane.
We call the solutions that satisfy this condition as the symmetric mode.
On the other hand, the condition $d\psi(0)/d\zeta=0$ allows gas to cross
the mid-plane.
We call the solutions that satisfy this second condition as the antisymmetric
mode.

We introduce an index, $i$, and define
\begin{equation}
\zeta_i \equiv (i-1)\Delta\zeta, \;\; \Delta\zeta
\equiv \frac{\zeta_{\rm node}}{N-1},
\end{equation}
where $N$ is the total number of points in the interval,
[0, $\zeta_{\rm node}$].
With a central difference scheme the finite-difference representation of 
the perturbation equation (\ref{ap:master}) is
\begin{equation}
\psi_{i-1} + [\Delta\zeta^2(E-V_o\zeta_i^2)-2]\psi_i + \psi_{i+1} = 0,
\label{fdeq}
\end{equation}
where $i$ runs from 3 to $N-2$ for the symmetric mode and from 2 to $N-2$
for the antisymmetric mode.
Because $\psi_{N} = 0$ at the upper boundary, the finite-difference
equation at the point $i=N-1$ is 
\begin{equation}
\psi_{N-2} + [\Delta\zeta^2(E-V_o\zeta_{N-1}^2)-2]\psi_{N-1} = 0.
\end{equation}
For the symmetric mode with $\psi_{1}=0$, we use the finite-difference
equation at the point of $i=2$,
\begin{equation}
[\Delta\zeta^2(E-V_o\zeta_2^2)-2]\psi_2 + \psi_{3} = 0.
\end{equation}
For the antisymmetric mode with $\psi_0 = \psi_2$, we use the
finite-difference equation at the point of $i=1$,
\begin{equation}
[\Delta\zeta^2(E-V_o\zeta_1^2)-2]\psi_1 + 2\psi_{2} = 0.
\end{equation}

We can write the simultaneous equations for $\psi_i$ in a matrix form,
\begin{equation}
M\psi = \left(
              \begin{array}{ccccc}
                     * & * &      &   &    \\
                    ** & * &  *   &   &    \\
                       &   &\ddots&   &    \\
                       &   &  *   & * & *  \\
                       &   &      & * & * 
              \end{array}
        \right)
        \left( 
              \begin{array}{c}
                  \psi_s \\ \psi_{s+1} \\ \vdots \\ \psi_{N-2} \\ \psi_{N-1}
               \end{array}
        \right) 
      = 
        \left(
              \begin{array}{c}
                     0 \\ 0 \\ \vdots \\ 0 \\ 0
              \end{array}
        \right),
\end{equation}
where $s$ is 2 for the symmetric mode and 1 for the antisymmetric mode.
Non-trivial solutions exist, only if the determinant of the tri-diagonal
matrix $M$ is equal to zero.
For given $\alpha$, $\beta$, and $\gamma$, the coefficients of the matrix
$M$ are functions of $\nu_x^2$, $\nu_y^2$, and $\Omega^2$.  
So, the problem becomes an eigenvalue problem, \ie, with given $\nu_x^2$
and $\nu_y^2$, we should find $\Omega^2$ which makes the determinant of
$M$ equal to zero.
We used the Gauss-Seidel method for it.

\clearpage

\figcaption{By the signs of $E$ and $V_o$, the solutions of the
eigenvalue equation (\ref{master}) are classified into four types.
For Type I and II, the solutions belong to the continuum family.
For Type III the solutions belong to the discrete family.
For Type IV, the solutions do not satisfy the imposed boundary condition.
\label{fig1}}

\figcaption{(a) Division of the phase space of $\nu_y^2$ and $\Omega^2$
into regions with the solutions of different types for a
non-magnetized disk with $\gamma=1$.  
The dashed line is the locus of $E=0$.
(b) Dispersion relations of the discrete family solutions with $n=0,~1,~2,
\cdots,6$ in the region of Type III.
\label{g1}}

\figcaption{(a) Division of the phase space of $\nu_y^2$ and $\Omega^2$
into regions with the solutions of different types for a
non-magnetized disk with $\gamma=0.8$.  
Three loci of $E=0$ (the dashed line), $V_o=0$ (the dotted line), and 
$\Omega^2=0$ (the long-dashed line) are drawn.
(b) Dispersion relations of the discrete family solutions with $n=0,~1,~2,
\cdots,6$ in the region of Type III and the dispersion relations of the
continuum family solutions with nodal points $\zeta_{\rm node}=3,~5,~10$
in the region of Type II.
\label{g0.8}}

\figcaption{(a) Division of the phase space of $\nu_y^2$ and $\Omega^2$
into regions with the solutions of different types for a
non-magnetized disk with $\gamma=1.3$.  
Three loci of $E=0$ (the dashed line), $V_o=0$ (the dotted line), and 
$\Omega^2=0$ (the long-dashed line) are drawn.
(b) Dispersion relations of the discrete family solutions with $n=0,~1,~2,
\cdots,6$ in the region of Type III and the dispersion relations of the
continuum family solutions with nodal points $\zeta_{\rm node}=3,~5,~10$
in the region of Type II.
\label{g1.3}}

\figcaption{Dispersion relations of the symmetric mode solutions (solid lines)
and the antisymmetric mode solutions (dotted lines) of the continuum family
with nodal points $\zeta_{\rm node}=2$ and $3$ for a non-magnetized disk with
$\gamma=0.8$.
\label{BC}}

\figcaption{(a) Division of the phase space of $\nu_y^2$ and $\Omega^2$ into
regions with the solutions of different types for a magnetized disk
with $\alpha=1$, $\beta=0$, $\gamma=1$, and $\nu_x=0$.
The loci of $E=0$ (the two dashed lines), $V_o=0$ (the dotted line), and 
$(2\alpha+\gamma)\Omega^2-2\alpha\gamma\nu_y^2=0$ (the long-dashed line)
are drawn.
(b) Dispersion relations of the discrete family solutions with
$n=0,~1,~2$ in the regions of Type III.  The dispersion relations of
the continuum family solutions with nodal points $\zeta_{\rm node}=3,~5,~10$
in the regions of Type I and II.
\label{fig:a1g1}}

\figcaption{(a) Division of the phase space of $\nu_y^2$ and $\Omega^2$ into
regions with the solutions of different types for a magnetized disk
with $\alpha=1$, $\beta=0$, $\gamma=\gamma_{\rm cr}$, and $\nu_x=0$.
The loci of $E=0$ (the two dashed lines), $V_o=0$ (the dotted line), and 
$(2\alpha+\gamma)\Omega^2-2\alpha\gamma\nu_y^2=0$ (the long-dashed line)
are drawn.
(b) Dispersion relations of the discrete family solutions with
$n=0,~1,~2$ in the regions of Type III and the dispersion relations of
the continuum family solutions with nodal points $\zeta_{\rm node}=3,~5,~10$
in the regions of Type I and II.
(c) Magnification of the region of Type III enclosed by loci of $E^-=0$
and $V_o=0$.
(d) Dispersion relations of the discrete family solutions with $n=0$,
1 and 2 in the magnified region of Type III.
\label{fig:a1gcr}}

\figcaption{Eigenfunctions of the unstable discrete mode solutions
with $n=0,~1,~2$ for $\alpha=1$, $\beta=0$, $\gamma=\gamma_{\rm cr}$, 
and $\nu_x=0$. For a given $n$, the most unstable growth rate
and its azimuthal wavenumber are used.
\label{fig8}}

\figcaption{Contours of equi-$\nu_{y,\rm int}^2$ (a) and 
equi-$\Omega_{\rm int}^2$
(b) for $\beta=0$ on the $(\alpha,~\gamma)$ plane.
The point ($\nu_{y,\rm int}^2$, $\Omega_{\rm int}^2$) is defined by the
intersection of the two loci of $E^-=0$ and $V_o=0$.
The intersection point exists only if $\gamma$ has a
value in the region enclosed by the lines labeled with $0$ and $\infty$.
\label{fig:intpoint}}

\figcaption{(a) Division of the phase space of $\nu_y^2$ and $\Omega^2$ into
regions with the solutions of different types for a magnetized disk
with $\alpha=1$, $\beta=0$, $\gamma=1$, and $\nu_x^2=1$.
The loci of $E=0$ (the two dashed lines), $V_o=0$ (two dotted lines),
$(2\alpha+\gamma)\Omega^2-2\alpha\gamma\nu_y^2=0$ (the long-dashed line),
and $\Omega^2-2\alpha\nu_y^2=0$ (the dot-dashed line) are drawn.
(b) Dispersion relations of the discrete family solutions with
$n=0,~1,~2$ in the regions of Type III and the dispersion relations of
the continuum family solutions with nodal points $\zeta_{\rm node}=3,~5,~10$
in the regions of Type I and II.
\label{fig:general}}


\clearpage

\begin{thebibliography}{}

\bibitem[Abramowitz \& Stegun 1970]{abr70} 
Abramowitz, M., \& Stegun, I. A. 1970, Handbook of Mathematical Functions 
(New York: Dover), 685.
\bibitem[Alfaro \etal 1992]{alf92} 
Alfaro, E. J., Cabrera-Ca\~{n}o, J., \& Delgado, A. J. 1992, \apj, 399, 576.
\bibitem[Appenzeller 1974]{app74} 
Appenzeller, I. 1974, \aap, 36, 99.
\bibitem[Begelman \& Zweibel 1994]{bz94}
Begelman, M. C., \& Zweibel, E. G. 1994, \apj, 431, 689.
\bibitem[Blitz \& Shu 1980]{bli80} 
Blitz, L., \& Shu, F. H. 1980, \apj, 238, 148.
\bibitem[Drury \& Falle 1986]{df86}
Drury, L. O'C., \& Falle, S. A. E. G. 1986, \mnras, 223, 353.
\bibitem[Drury \& V\"olk 1981]{dv81}
Drury, L. O'C., \& V\"olk, H. J. 1981, \apj, 248, 344.
\bibitem[Elmegreen 1982]{elm82}
Elmegreen, B. G. 1982, \apj, 253, 634.
\bibitem[Falgarone \& Lequeux 1973]{fl73}
Falgarone, E., \& Lequeux, J. 1973, \aap, 25, 253.
\bibitem[Foglizzo \& Tagger 1994]{fog94} 
Foglizzo, T., \& Tagger, M. 1994, \aap, 287, 297.
\bibitem[Florido \etal 1991]{flo91} 
Florido, E., Battaner, E., Prieto, M., Mediavilla, E., \& Sanchez-Saavedra,
M.  L. 1991, \mnras, 251, 193.
\bibitem[Giz \& Shu 1993]{giz93} 
Giz, A. T., \& Shu, F. H. 1993, \apj, 404, 185.
\bibitem[Gomez de Castro \& Pudritz 1992]{gom92} 
Gomez de Castro, A. I., \& Pudritz, R. E. 1992, \apj, 395, 501.
\bibitem[Hanawa \etal 1992]{han92} 
Hanawa, T., Matsumoto, R., \& Shibata, K. 1992, \apj, 393, L71.
\bibitem[Horiuchi \etal 1988]{hor88}
Horiuchi, T., Matsumoto, R., Hanawa, T., \& Shibata, K. 1988, \pasj, 40, 147.
\bibitem[Kaisig \etal 1990]{kai90}
Kaisig, M., Tajima, T., Shibata, K., Nozawa, S., \& Matsumoto, R.
1990, \apj, 358, 698.
\bibitem[Kang, Jones, \& Ryu 1992]{kjr92}
Kang, H., Jones, T. W., \& Ryu, D. 1992, \apj, 385, 193.
\bibitem[Lockman 1977]{loc77} 
Lockman, F. J. 1977, \aj, 82, 408.
\bibitem[Malhotra 1994]{mal94} 
Malhotra, S. 1994, \apj, 433, 687.
\bibitem[Matsumoto \etal 1988]{mat88} 
Matsumoto, R., Horiuchi, T., Shibata, K., \& Hanawa, T. 1988, \pasj, 40, 171.
\bibitem[Mouschovias 1974]{mou74a} 
Mouschovias, T. Ch. 1974, \apj, 192, 37.
\bibitem[Mouschovias \etal 1974]{mou74b} 
Mouschovias, T. Ch, Shu, F. H., \& Woodward P. R. 1974, \aap, 33, 73.
\bibitem[Nelson 1976]{nel76} 
Nelson, A. H. 1976, \mnras, 174, 661.
\bibitem[Nozawa \etal 1992]{noz91}
Nozawa, S., Shibata, K., Matsumoto,  R., Sterling, A. C.,  Tajima, T.,
Uchida, Y.,  Ferrari, A., \& Rosner, R. 1992, \apjs, 78, 267.
\bibitem[Oort 1965]{oor65} 
Oort, J. H.  1965, in Galactic  Structure, ed. A.  Blaauw, \&  M. Schmidt
(Chicago:  Univ. of Chicago Press), 455.
\bibitem[Parker 1966]{par66} 
Parker, E. N. 1966, \apj, 145, 811.
\bibitem[Parker 1967]{par67}
Parker, E. N. 1967, \apj, 149, 535.
\bibitem[Quiroga \& Schlosser 1977]{qui77} 
Quiroga, R. J., \& Schlosser, W. 1977, \aap, 57, 455.
\bibitem[Ryu \& Goodman 1994]{ryu94} 
Ryu, D., \& Goodman, J. 1992, \apj, 388, 438.
\bibitem[Ryu, Kang, \& Jones 1993]{rkj93}
Ryu, D., Kang, H., \& Jones, T. W. 1993, \apj, 405, 199.
\bibitem[Shibata \& Matsumoto 1991]{shi91} 
Shibata, K., \& Matsumoto, R. 1991, \nat, 353, 633.
\bibitem[Shibata \etal 1989a]{shi89a}
Shibata, K., Tajima, T., Matsumoto, R., Horiuchi, T., Hanawa, T.,
Rosner, R., \& Uchida, Y. 1989a, \apj, 338, 471.
\bibitem[Shibata \etal 1989b]{shi89b}
Shibata, K., Tajima, T., Steinolfson, R. S., \& Matsumoto, R. 
1989b, \apj, 345, 584.
\bibitem[Shu 1974]{shu74} 
Shu, F. H. 1974, \aap, 33, 55.
\bibitem[Skilling 1975]{Skilling75}
Skilling, J. 1975, \mnras, 172, 557.
\bibitem[Spicker \& Feitzinger]{spi86}
Spicker, J., \& Feitzinger, J. V. 1986, \aap, 163, 43.
\bibitem[Spitzer 1978]{spi78}
Spitzer, L., Jr. 1978, Physical Processes in the Interstellar Medium
(New York: John Wiley \& Sons), 226.
\bibitem[Tout \& Pringle 1992]{tou92}
Tout, C. A., \& Pringle, J. E. 1992, \mnras, 259, 604.
\bibitem[Zank, Axford, \& McKenzie 1990]{zam90}
Zank, G. P., Axford, W. I., \& McKenzie, J. F. 1990, \aap, 233, 275.
\bibitem[Zweibel \& Kulsrud 1975]{zei75} 
Zweibel, E. G., \& Kulsrud, R. M. 1975, \apj, 201, 63.

\end{thebibliography}
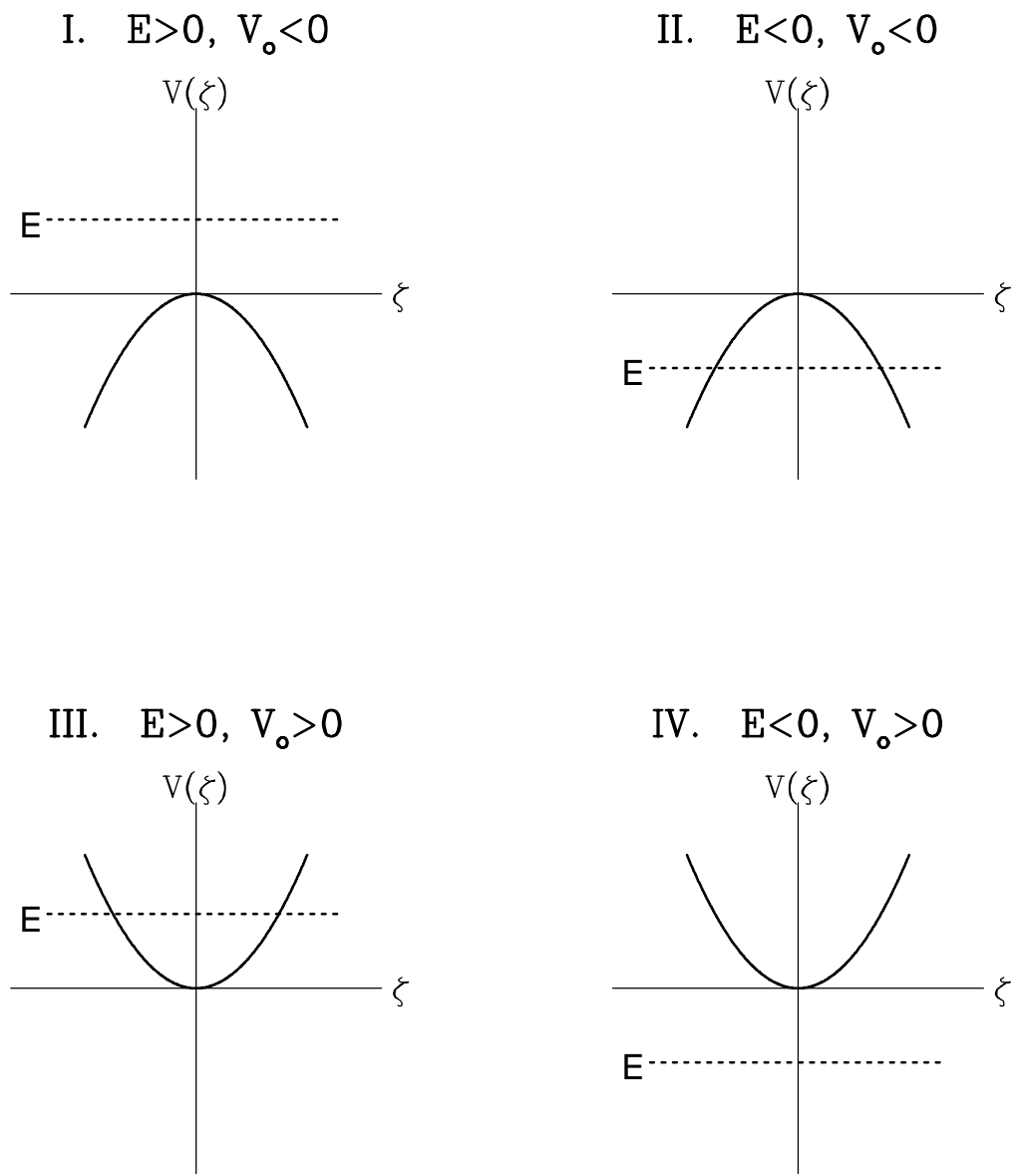
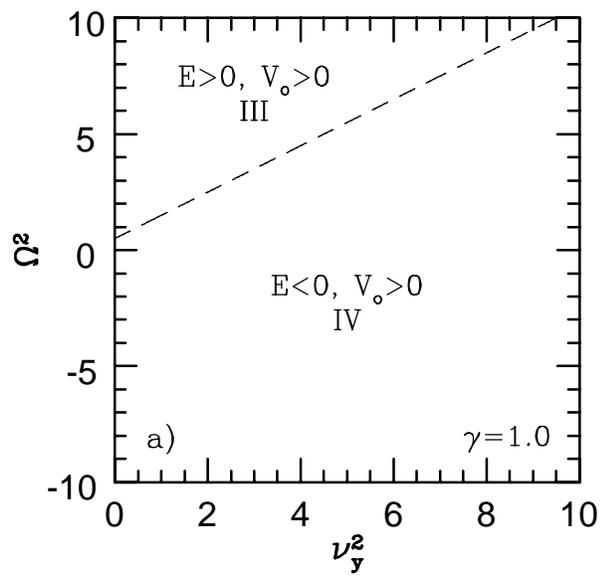
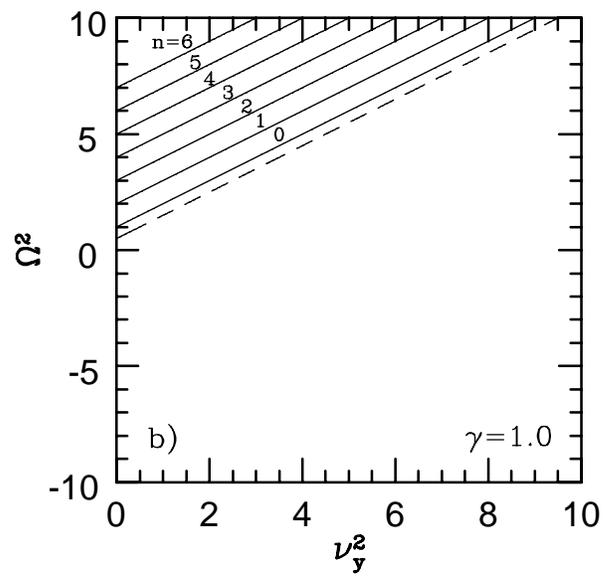
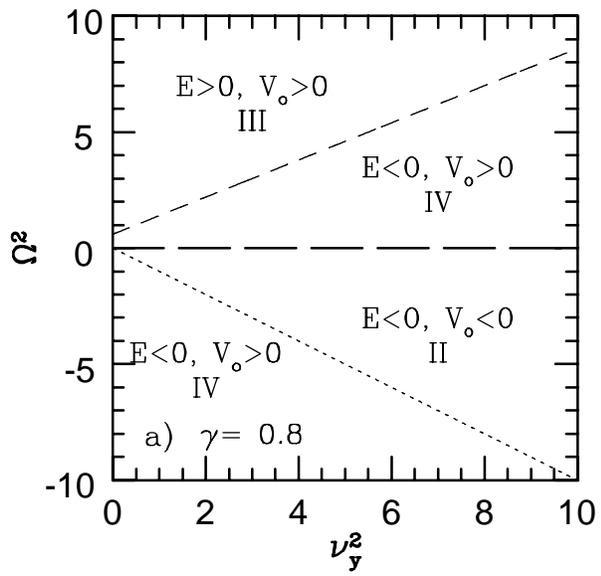
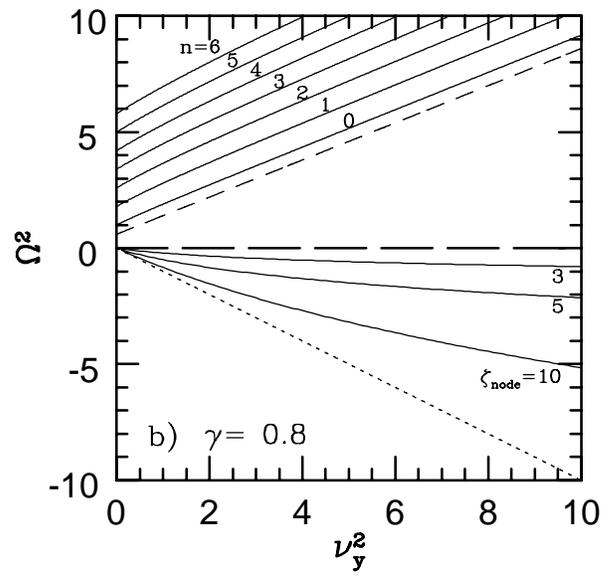
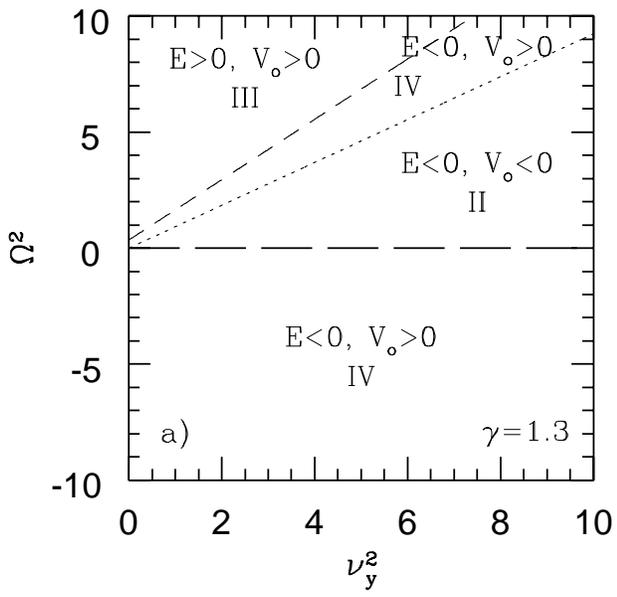
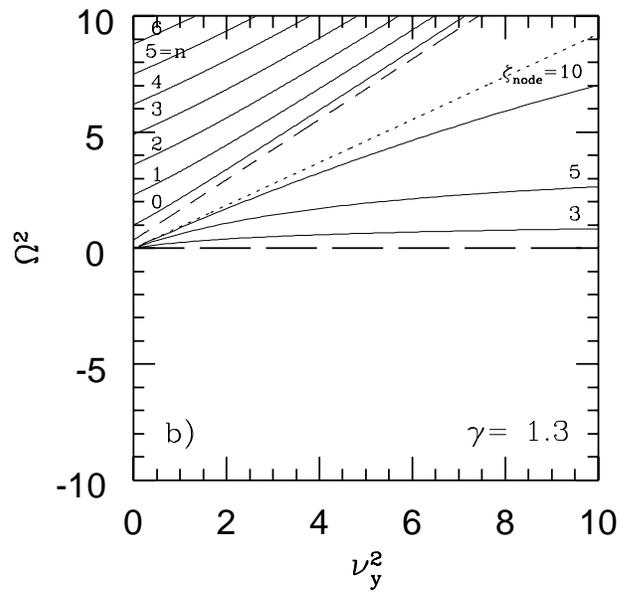
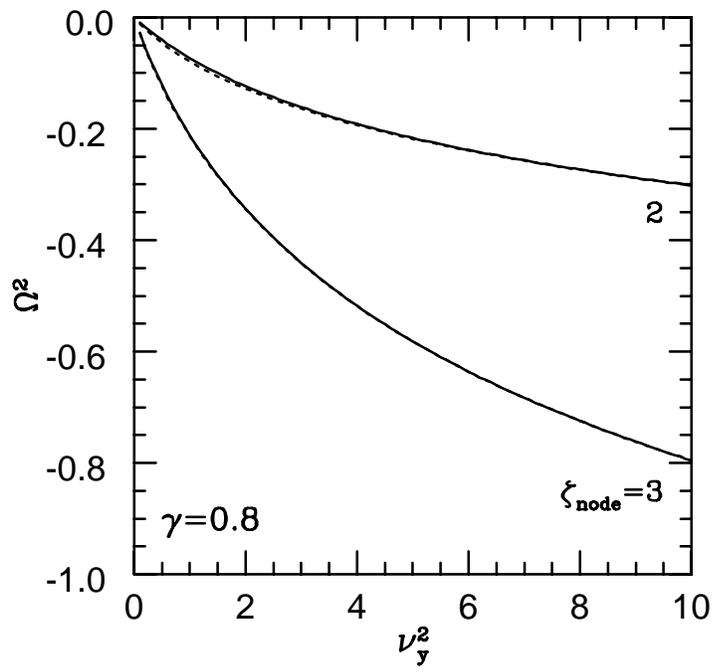
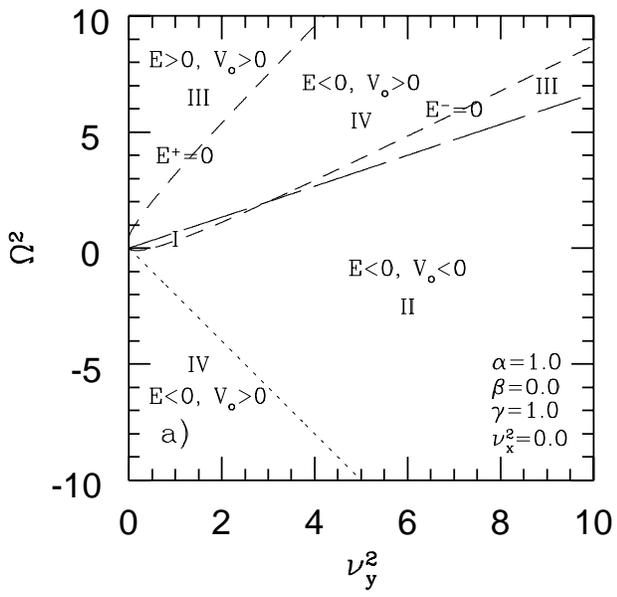
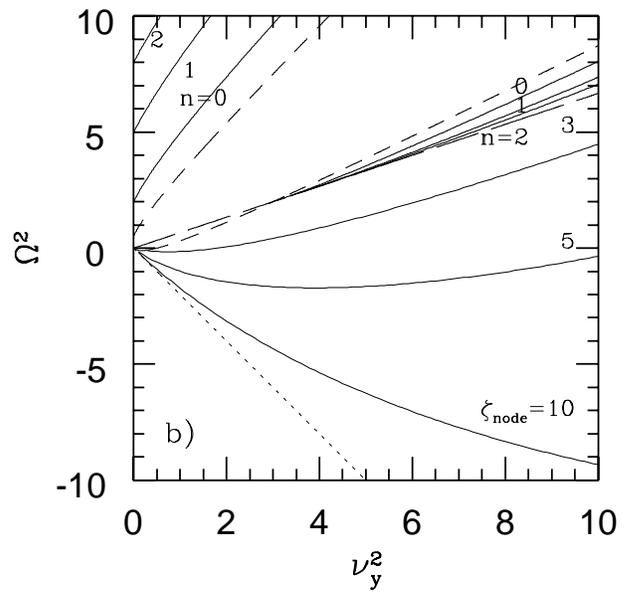
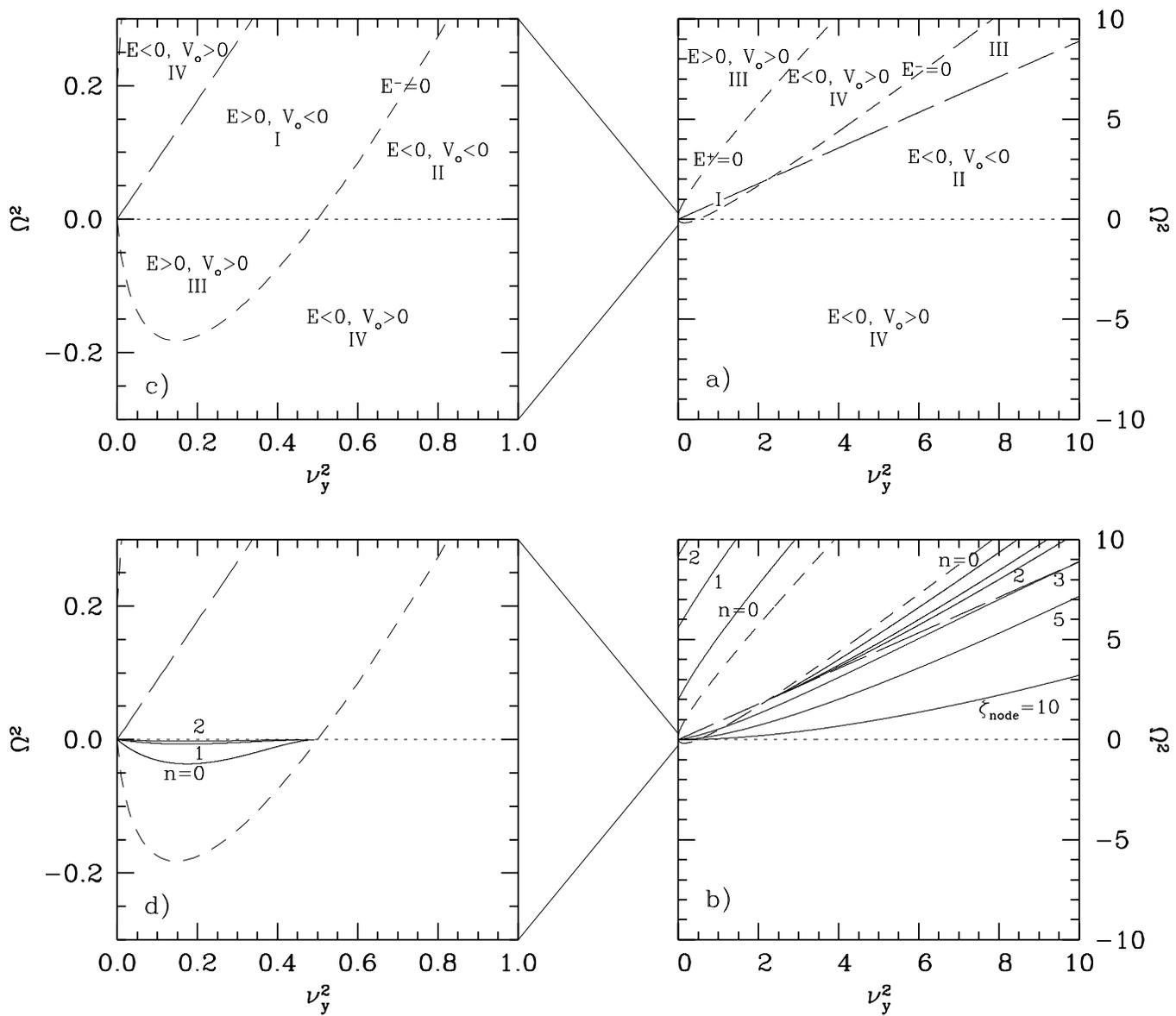
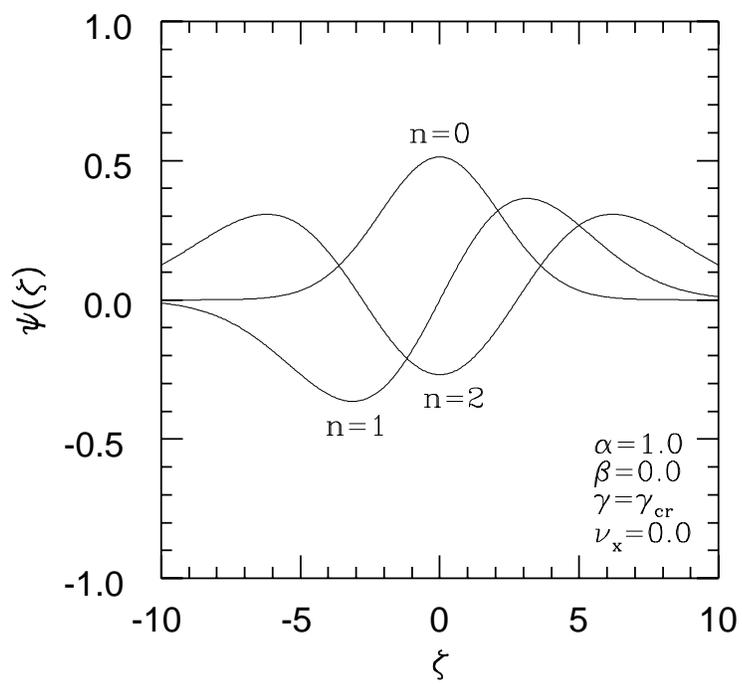
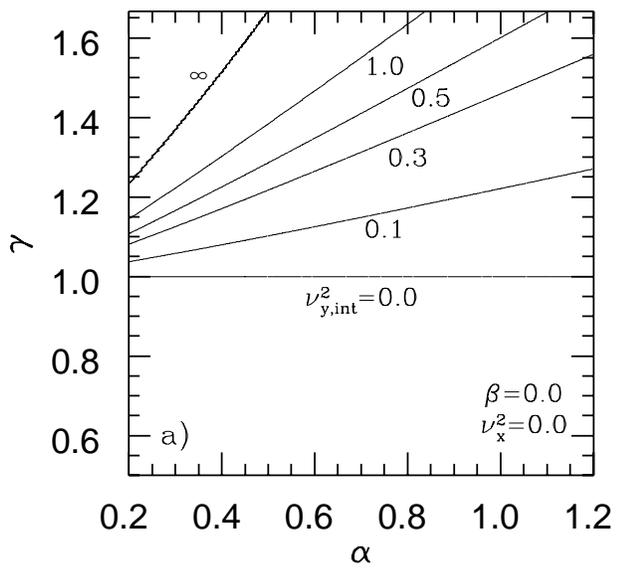
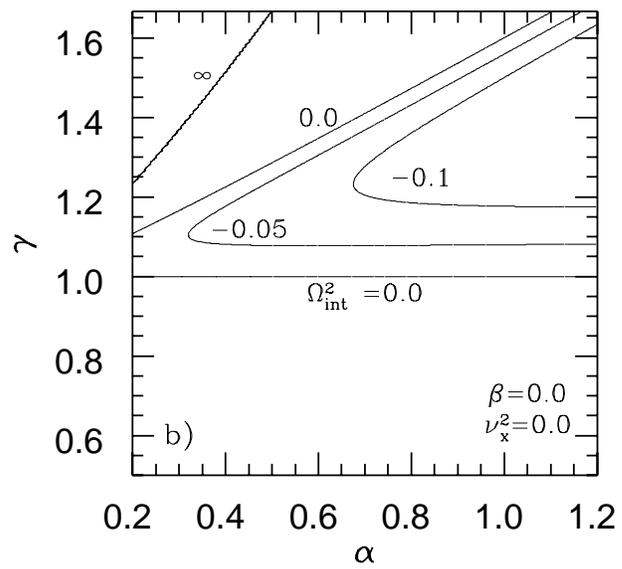
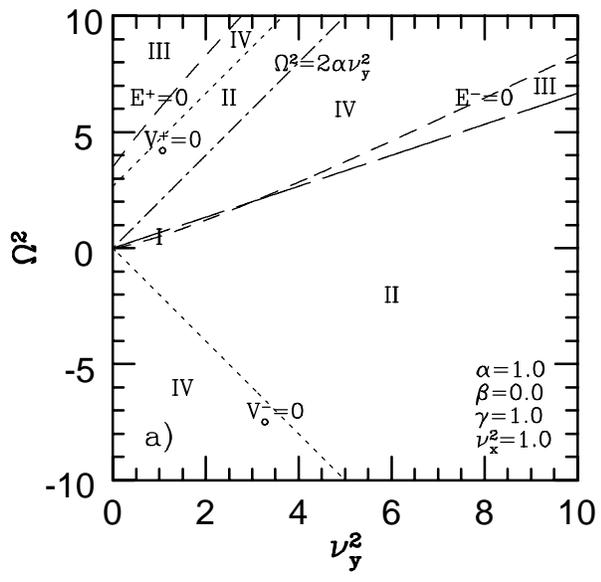
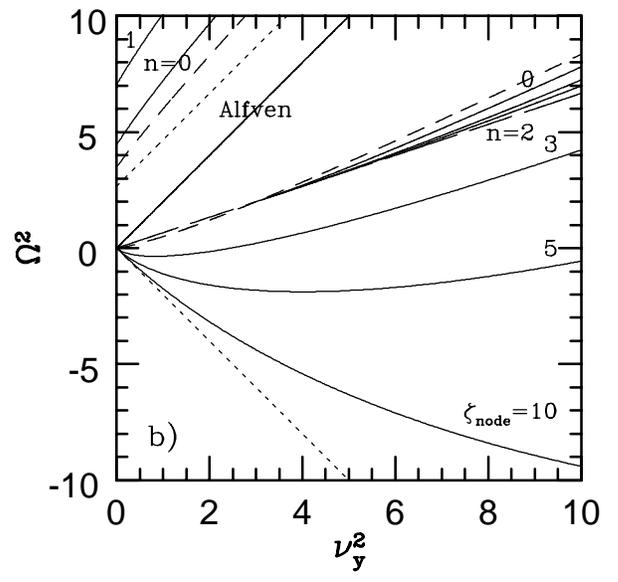
\end{document}